\documentclass[aps,pre,showpacs,superscriptaddress,twocolumn]{revtex4}
\usepackage{graphicx,psfrag}

\begin{document}
\title{Front motion in an $A+B\rightarrow C$ type reaction-diffusion process: \\
Effects of an electric field}

\author{I. Bena}
\affiliation{Department of Physics, 
University of Gen\`eve, CH-1211 Gen\`eve 4, Switzerland}
\author{F. Coppex}
\affiliation{Department of Physics, 
University of Gen\`eve, CH-1211 Gen\`eve 4, Switzerland}
\author{M. Droz}
\affiliation{Department of Physics,
University of Gen\`eve, CH-1211 Gen\`eve 4, Switzerland}
\author{Z. R\'acz}
\affiliation{Institute for Theoretical Physics - 
HAS, E\"otv\"os University, P\'azm\'any s\'et\'any 1/a, 1117 Budapest, Hungary}

\date{\today}

\begin{abstract}

We study the effects of an external electric field on both
the motion of the reaction zone and the spatial distribution
of the reaction product, $C$, in an irreversible
$A^- +B^+ \to C$ reaction-diffusion
process. The electrolytes $A\equiv (A^+,A^-)$ and
$B\equiv (B^+,B^-)$ are initially separated in space and
the ion-dynamics is described by reaction-diffusion equations
obeying local electroneutrality. Without an electric
field, the reaction zone moves diffusively leaving behind
a constant concentration of $C$-s. In the presence
of an electric field which drives the reagents towards the
reaction zone, we find that the reaction zone still moves
diffusively but
with a diffusion coefficient which slightly decreases with increasing
field. The important electric field effect is that the
concentration of $C$-s is no longer constant but increases linearly
in the direction of the motion of the front.
The case of an electric field of
reversed polarity is also discussed and it 
is found that the motion of the front
has a diffusive, as well as a drift component.
The concentration of $C$-s decreases
in the direction of the motion of
the front, up to the complete extinction of the reaction. 
Possible applications of the above results
to the understanding of the formation of Liesegang patterns
in an electric field is briefly outlined.

\end{abstract}
\pacs{05.60.Gg, 64.60.Ht, 75.10.Jm, 72.25.-b}
\maketitle

\section{Introduction}
\label{introduction}

Nonequilibrium systems exhibiting pattern formation
are ubiquitous in nature
and these patterns often emerge in the wake
of a moving reaction front~\cite{cross94}.
A classical example which motivated our present study
is the emergence of {\em Liesegang bands}~\cite{liesegang1896,henisch91}
associated with the generic reaction scheme $A+B\rightarrow C$
and the subsequent precipitation of $C$-s.
In a typical experimental setup, a chemical reactant, $B$
(called inner electrolyte), is disolved in a gel matrix,
while a second reactant, $A$
(outer electrolyte), of much higher concentration is brought
into contact with the gel.
The outer electrolyte diffuses into the gel, reacts with
the inner electrolyte, and the reaction front moves
into the gel.
Under appropriate conditions, the reaction product precipitates
and, depending on the geometry of the experimental setup,
one can observe the emergence of families of bands or rings
perpendicular to the direction of motion of the 
front~\cite{henisch91,Muller-Ross}.

This pattern formation process is a rather complex one, due
to the delicate interplay between the
motion of the reaction front and the precipitation dynamics 
of the reaction product $C$. 
Its study has a history of more than a century~\cite{liesegang1896}
and the generic, empirical laws describing it are
well established~\cite{racz99}.
There is still disagreement, however, about the mechanisms
underlying this pattern-forming process
(see~\cite{antal98,magnin00} for a comparative analysis of
the existing theories). Moreover, with the exception of the
recently-proposed phase separation (or
spinodal decomposition) scenario~\cite{antal99,antal01},
the existing theories 
are essentially {\em qualitative}, as they contain 
parameters that cannot be inferred
from experiments, and therefore they do not allow for
{\em quantitative} predictions or for any direct
comparison with the existing experimental data.
Having this state of affairs, we believe that
it is important to test the spinodal decomposition
theory in as many ways as possible.

The basic ingredients of the spinodal decomposition approach
are the presence of a moving reaction front and the phase
separation that takes place in the wake of the front.
The dynamics of the {\em inert} reaction product
$C$ is assumed to have no feedback on the dynamics
of the reagents~\cite{footnote1}, thus one represents the dynamics
as a {\em two-stage process}. Namely,
(i) the production of $C$ in 
the moving reaction front is described by reaction-diffusion 
equations, while (ii) the diffusion and precipitation
of $C$ particles is modelled as
phase separation (spinodal decomposition)~\cite{gunton83}
using the Cahn-Hilliard equation~\cite{cahn58,cahn61,hohenberg77},
with a source term corresponding to the
production of $C$ by stage (i). All the parameters involved in this model
(including the parameters of the Landau-Ginzburg 
free energy associated with the dynamics of $C$ particles)
are either directly accessible experimentally or can be inferred
from experimental data (see~\cite{racz99}).

The first stage $(A+B\rightarrow C)$ of the process
appears in many other physical and chemical processes and has been
much studied. The
dynamics of the front and the spatial distribution of 
the rate of the
production of $C$ (the source for the second stage) are known~\cite{galfi88}
for the case of neutral reagents $A$ and $B$. It should be
realized, however, that $A$ and $B$ are usually electrolytes
which dissociate,
\begin{equation}
A \rightarrow A^++ A^-\,,\qquad B \rightarrow B^++B^-\,,
\label{dissociation}
\end{equation}  and the basic reaction process takes place between 
the `active' ions, e.g., 
\begin{equation}
A^-+B^+\rightarrow C \,,
\label{reaction}
\end{equation}
while the `background' ions $A^+$ and $B^-$ are not reacting.
For the first sight, the background ions may be not important.
They just ensure {\em local electroneutrality}
and the Debye screening length is much smaller than the
spatial lengths involved in the macroscopic pattern formation.
However, as it was shown recently~\cite{unger00}, the
dynamics of the background ions may generate macroscopic effects
(even if the screening length is negligible) and may influence
both the propagation of the reaction front
and the structure of the resulting pattern.
Thus, an obvious way to obtain more insight into the
details of the dynamics of Liesegang pattern formation
(with the ultimate goal of validating one theoretical
scenario or another) is to switch on an {\em electric field}
and to study its effects.

Experimental and theoretical investigations in this direction
have been going on for quite a while, 
see~\cite{happel29,kisch29,ortoleva78,feeney83,sharbaugh89,das90,das91,das92,muller91,muller94,sultan00,ghoul03,lagzi02,lagzi03}. 
The emerging picture, however, might be sometimes confusing.
For example, on the experimental side, Das et al.'s 
experiments~\cite{das90,das91}
show a diffusive motion of the front for a polarity of 
the applied field that favors 
the reaction, and for a reverse polarity, 
as shown by Sultan et al.~\cite{sultan00} the
motion of the front acquires a supplementary drift component.
These results are just the opposite of those obtained by Lagzi~\cite{lagzi02},
however it can be argued~\cite{lagzi04} that in the latter case
the properties of the intermediate compounds are responsible for this
`anomalous' behavior.
As far as the theories
are concerned, a drift of the front is obtained analytically in 
e.g.~\cite{feeney83} under the irrealistic assumption of the constancy of the 
electric field along the system, or it is put `by hand' as e.g. 
in~\cite{lagzi03}.

In order to clarify the effect of the electric field on
Liesegang patterns, we propose to revisit the problem
in the framework of the spinodal decomposition theory.
In the present paper, we shall concentrate on the first 
stage of this scenario, namely on the effect of the 
electric field on the motion
of the reaction front and on the concentration of the
reaction product $C$ left behind the front. As mentioned above,
the problem of the first stage is more general and thus
we feel that the results are of interest in their own. Although
we shall mention the implications of the reaction zone
results for Liesegang patterns, the calculations and discussions
of the second stage of the process will be presented in
a forthcoming study.

As it will be discussed in the next section
(Sec.~\ref{theModel}, with some details in Appendix~A),
the new element in our model is that the dynamics of all
the ions $(A^+,A^-,B^+,B^-)$ is followed by using
reaction-diffusion equations obeying the
electroneutrality condition. Furthermore, the electric field
is taken into account by prescribing a given
potential drop between the two ends of the sample,
as in real experimental situations. The
equations are solved numerically in Sec.~\ref{simulations} where,
in view of the possible comparison of our results
with experimental data, we concentrate our study on
the {\em experimentally relevant} range of parameters and
observation times.

Our results can be summarized by
contrasting them with the fieldless case where the reaction
front moves diffusively, and it leaves behind
a constant concentration $c_0$ of the reaction product 
$C$~\cite{galfi88,unger00}. In the presence of the electric field,
the polarity of the field is an important factor.
If the field favours the reaction (in the
sense that it drives the reacting ions $A^-$ and $B^+$ towards the
reaction zone), then the motion of the reaction front
remains diffusive, and the concentration of the
reaction product $c(x)$ increases linearly in space in the direction
of the motion of the front. The diffusive front motion is in
agreement with Das et al.'s experiments~\cite{das90,das91} and
in contrast with theoretical conjectures about constant-velocity
drifts~\cite{feeney83,lagzi03}. 
An important remark is that this case
is difficult to handle, since a strong electric field develops in the reaction
zone (especially for large applied tensions), which leads to numerical
instabilities and thus restricts our numerical
solutions to relatively short times.

In the case of opposite polarity of the applied field, 
the motion of the reaction zone is still diffusive
at short times, but a crossover to drift at large times,
as seen in Sultan's experiments~\cite{sultan00},
can be inferred from numerical data.
As far as the concentration of the
reaction product is concerned, we find that at small times
it decreases linearly in the direction of
the motion of the front, while
a slow-down of this decrease is noticed for larger times,
up to the extinction of the reaction.

The conclusions and the
implications of the results for Liesegang phenomena
will be presented in Sec.~IV. Finally, some salient details of
the problem such as the electric-field profiles, the origin of the numerical
instabilities,
the time evolution of
the electric current, the finite-size effects, etc. are
discussed in Appendix~B.


\section{The model in the presence of an applied electric field}
\label{theModel}

\begin{figure}[htbp]
\begin{center}
\psfrag{-LA}{$-L_A$}
\psfrag{LB}{$L_B$}
\psfrag{o}{$0$}
\psfrag{ApAm}{$A \, (A^+,A^-)$}
\psfrag{BpBm}{$B \, (B^+,B^-)$}
\psfrag{xf(t)}{$x_f(t)$}
\psfrag{Clow}{low-density $C$}
\psfrag{Chigh}{high-density $C$}
\psfrag{x}{$x$}
\psfrag{U}{$U$}
\psfrag{+}{$+$}
\psfrag{-}{$-$}
\psfrag{j0}{$j_0(t)$}
\vspace{1cm}
\includegraphics[width=\columnwidth]{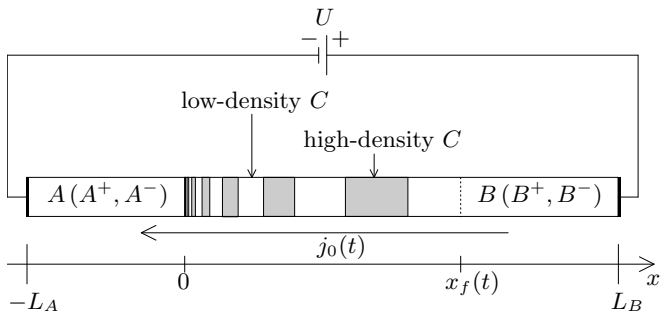}
\end{center}
\caption{Schematic representation of the system under study. 
The electrolytes $A\,(A^+,\,A^-)$
and $B\,(B^+,\,B^-)$ are initially located in the regions $[-L_A,\,0)$, 
respectively $(0,\,L_B]$.  
The precipitation bands, i.e., the alternation of high-density-$C$ regions 
(shaded areas) and low-density-$C$ ones, emerge in the wake of the 
moving reaction front (dashed line at $x_f(t)$).
For an applied tension $U$, an electric current of density $j_0(t)$ is 
flowing through the system.}\label{figure1}
\vspace{1cm}
\end{figure}

The experimental setup we shall consider is represented 
schematically in Fig.~\ref{figure1}. The two electrolytes are initially separated
from each other, each is uniformly spread into a gel matrix, and the concentration 
of $A$'s is much larger than that of $B$'s. 
Thus the reaction front $A^-+B^+\rightarrow C$ moves to the right
along the column. An appropriate choice of the experimental conditions
(type of reagents, concentrations, etc.) leads to quasiperiodic
precipitation of the inert reaction product $C$ in the wake of the moving front
-- the Liesegang patterns, see Fig.~\ref{figure1}. 
Due to the presence of the
gel the convection phenomena are absent, and thus the 
evolution of the concentration fields of the particles is accurately
described by {\em reaction-diffusion} equations.

Our model is based on several simplifying assumptions, and we refer
the reader to Appendix~A for  more details on a more general model.

(i) The first assumption is that the system is {\em one-dimensional},
which means that all the relevant quantities depend 
on a single spatial coordinate $x$, with
$-L_A\leq x \leq L_B$ being the spatial extent of the system.
This is essentially realized in practice provided 
that the length of the gel column is much larger than its width.
The edge effects in the transverse directions might have some 
slight consequences (e.g., as reported in \cite{das90},
in the presence of an external electric field
the diameter of the tube has a small influence on
the propagation of the reaction front). 
However, the final pattern is still one-dimensional 
to a very good accuracy.

(ii) We consider instantaneous {\em 100\% dissociation} of the electrolytes
$A$ and $B$ according to Eq.~(\ref{dissociation}) (the assumption of
``ideally strong" acid and basis). Correspondingly, only $A^{\pm}$ and $B^{\pm}$ ions
(and the reaction product $C$) are present in the system.

(iii) The ions $A^-$ and $B^+$ are reacting irreversibly with an {\em infinite
reaction rate}. This can be justified by the fact that the characteristic time of the 
reaction process is $3$ to $6$ orders of magnitude less than the time scales invloved
in the diffusion and precipitation processes, see, e.g., \cite{unger00,magnin00}.
In this case, one has a {\em pointlike reaction front}
at a time-dependent position $x_f(t)$ \cite{footnote2}.

(iv) The {\em electroneutrality} approximation~\cite{rubinstein90} 
-- according to which the local 
charge density is zero on space scales that 
are relevant to pattern formation -- reads 
\begin{equation}
\sum_i z_i \,n_i(x,t) =0\,,
\label{electroneutrality}
\end{equation}
where $z_i$ is the charge of the $i$-th ion -- in terms of the elementary charge $q$ -- and $n_i(x,t)$
denotes its density. As discussed in detail in~\cite{unger00,unger99}, this condition is well fulfilled
in the systems we are investigating. Moreover, we shall consider monovalent ions, $|z_i|=1$ for all $i$.

(v) Finally, we consider {\em equal diffusion coefficients} of the ions, $D_i=D$ for all $i$.

The initial conditions for the concentration 
profiles $a^{\pm}(x,t)$ and $b^{\pm}(x,t)$ 
of the $A^{\pm}$ and $B^{\pm}$ ions correspond to initially separated reagents,
\begin{eqnarray}
&&a^-(x,t=0)=a^+(x,t=0)={a_0}\,,\nonumber\\
&&b^+(x,t=0)=b^-(x,t=0)=0\,, \nonumber\\
&& \hspace{100pt}\mbox{for} \;\: -L_A\le x\le 0\, \nonumber\\
&&a^-(x,t=0)=a^+(x,t=0)=0 \,,\nonumber\\
&&b^+(x,t=0)=b^-(x,t=0)=b_0\,, \nonumber\\
&&\hspace{100pt} \mbox{for} \;\; 0< x\le L_B\,,
\label{initial_conditions}
\end{eqnarray}
with $a_0 \gg b_0$. We also suppose that the concentrations of the ions are maintained fixed at the borders,
\begin{eqnarray}
&&a^+(-L_A,t)=a^-(-L_A,t)={a_0}\,,\nonumber\\
&&b^+(-L_A,t)=b^-(-L_A,t)=0\,,\nonumber\\
&&a^+(L_B,t)=a^-(L_B,t)=0,\,\nonumber\\
&&b^+(L_B,t)=b^-(L_B,t)=b_0\,,
\end{eqnarray}
i.e., the system is in contact at its left and right borders 
with two `infinite' reservoirs of
ions $A^{\pm}$ and $B^{\pm}$, respectively.

Under these assumptions, the general 
evolution equations for the concentration profiles
$a^{\pm}(x,t)$ and $b^{\pm}(x,t)$ 
of the $A^{\pm}$ and $B^{\pm}$ ions
presented in the Appendix~A 
acquire the following 
simplified expressions:

(i) For  $x<x_f(t)$ (i.e., in the wake of the reaction front)
\begin{eqnarray}
&&\frac{\partial a^-(x,t)}{\partial t}=D\frac{\partial ^2 a^-}{\partial x^2} +\frac{j_0}{2q}
\frac{\partial}{\partial x}\left(\frac{a^-}{a^+}\right)\,,
\label{equations_left}\\
&&b^+(x,t)\equiv 0\,,
\\
&&\frac{\partial a^+(x,t)}{\partial t}=D \frac{\partial^2 a^+}{\partial x^2}\,,
\\
&&\frac{\partial b^-(x,t)}{\partial t}=D\frac{\partial ^2 b^-}{\partial x^2} +\frac{j_0}{2q}
\frac{\partial}{\partial x}\left(\frac{b^-}{a^+}\right) \,.
\end{eqnarray}

(ii) For $x>x_f(t)$ (i.e., ahead the reaction front)
\begin{eqnarray}
&&a^-(x,t)\equiv 0\,,
\\
&&\frac{\partial b^+(x,t)}{\partial t}=D\frac{\partial ^2 b^+}{\partial x^2} -\frac{j_0}{2q}
\frac{\partial}{\partial x}\left(\frac{b^+}{b^-}\right)\,,
\\
&&\frac{\partial a^+(x,t)}{\partial t}=D \frac{\partial^2 a^+}{\partial x^2} -\frac{j_0}{2q}
\frac{\partial}{\partial x}\left(\frac{a^+}{b^-}\right)\,,
\\
&&\frac{\partial b^-(x,t)}{\partial t}=D\frac{\partial ^2 b^-}{\partial x^2} \,.
\label{equations_right}
\end{eqnarray}

Here $j_0$ is the time-dependent electric current density
\begin{equation}
j_0(t)=\frac{-2qFDU}{RT\left(\displaystyle\int_{-L_A}^{x_f(t)}\frac{dx}{a^{+}}+
\displaystyle\int_{x_f(t)}^{L_B}\frac{dx}{b^-}\right)}
\label{current_density}
\end{equation}
determined by the {\em constant voltage difference} $U=V(L_B)-V(-L_A)$ applied
between the two ends of the system \cite{footnote3}. $F=qN_A$ 
is Faraday's constant (i.e., the electric charge
transported by a mole of monovalent positive ions), 
$R$ is the universal gas constant, while $T$ is the temperature.
Finally, the local electric field is given by:
\begin{eqnarray}
E(x,t)&=&\frac{j_0(t)}{\displaystyle\frac{qFD}{RT}\,\sum_in_i(x,t)}\nonumber\\
&=&\frac{-2U}{\displaystyle\sum_in_i
\left(\displaystyle\int_{-L_A}^{x_f(t)}\frac{dx}{a^{+}}+
\displaystyle\int_{x_f(t)}^{L_B}\frac{dx}{b^-}\right)}\,.
\label{electric_field}
\end{eqnarray}

At a first sight, one would be tempted to conclude that the
equations~(\ref{equations_left})--(\ref{equations_right})
for the concentration fields look simple,
but actually  (i) they are nonlinearly coupled through
the integral quantity $j_0(t)$, according to 
Eq.~(\ref{current_density}), and (ii)
they acquire {\em time-dependent boundary conditions},
\begin{eqnarray}
&&a^-(x_f(t))=b^+(x_f(t))=0\,,\nonumber\\
&&|j_{a^{-}}(x_f(t))|=|j_{b^{+}}(x_f(t))|\,.
\end{eqnarray}
which are complicated to handle. The
meaning of these latter conditions is just that 
the concentrations of the reagents
are zero at the front, and their  flux towards the reaction front are equal.

Even under the simplifying assumption presented above,
an analytic solution for the concentration profiles
of the ions and thus for the motion of the front 
seems hopeless. We thus turned to numerical solutions,
whose main results are presented in the following Section. 


\section{Numerical results}
\label{simulations}

As already mentioned in the Introduction, we decided to focus our analysis 
on situations that are experimentally relevant. In particular, although we are aware
of different other regimes that might appear, we restrict the presentation
to the ones that are most likely to appear in experiments.
The choice of the parameters intends thus to mimic 
real experimental situations, namely, we considered concentrations of the ions
$a_0$ and $b_0$ of the order $10^{-3}$ M -- $10$ M, lengths of the system 
$L=L_{A}+L_B$
of some tenths centimeters, and tensions $U$ applied 
between system's edges 
such that the corresponding field intensity $U/L$ varies between
$\pm 15$ V/m. The common
diffusion coefficient of the ions was chosen as 
$D=10^{-9}$ m$^2/$s for all the
calculations.
Finally, times of observation of the system 
that vary between a few hours and some tenth days. 
Note that this observation time corresponds
to the real experimental time and {\em not} to the time
required by the numerical calculations. As far as the
actual numerical calculations are concerned,
we integrated the partial differential equations
(\ref{equations_left})--(\ref{equations_right})
using a fourth-order Runge-Kutta integration scheme in time and 
a first-order differenciation procedure in space.

We shall generally present the results that were obtained for 
a polarity $U>0$ of the applied field that favors the reaction,
i.e., that drives the $A^-$ and $B^+$
ions towards the reaction zone.
This is the case that is apriori more
interesting for the formation of the
Liesegang pattern.
However, this situation is numerically difficult, 
since a strong electric field develops rapidly (especially 
for large applied tensions) in the reaction zone, 
thus leading to the appearance
of numerical instabilities which limit the observation time.
For more details on this point see Appendix~B.

We also indicate the main results for the opposite 
polarity of the electric field, 
for which the production of $C$ is slowed down and finally stopped.
This is a situation that becomes rather rapidly
uninteresting from the point of view of the pattern
formation.

\subsection{Front propagation}
\label{front}

In the limit of an infinite reaction rate, the reaction zone
reduces to a point whose position $x_f(t)$ is moving
to the right in the experimental setup of Fig.~\ref{figure1}.

The main result of  the simulations we carried on for the case of
the field favoring the reaction is that the motion of the
front is {\em diffusive} with an effective diffusion
coefficient $D_f$,
\begin{equation}
x_f(t)=\sqrt{2D_f t}\,.
\end{equation}  
The above statement is valid
as far as the border effects are negligible (see below)
and for the parameter domains that are experimentally relevant. 
This result is in agreement with the rather accurate
experimental results of Das et al. \cite{das90,das91},
but contradicts some previous theoretical considerations
\cite{feeney83,lagzi03}
which, however, in our opinion, have no  justification.

\begin{figure}[h!]
\begin{center}
\vspace{1cm}
\includegraphics[width=\columnwidth]{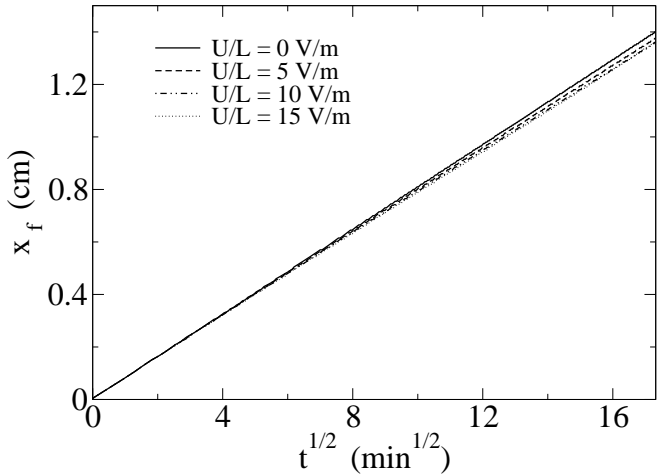}
\end{center}
\caption{Position of the reaction front as a function of the square root of
time for different
values of the intensity $U/L$ of the applied field.
The values of the other parameters
are: $L_A=1$  cm, 
$L_B=20$ cm, $a_0=10$ M, and $b_0=0.1$ M.
The observation time is limited to $5$ hours (see the main text).}
\label{figure2}
\vspace{1cm}
\end{figure}

Figure~\ref{figure2} shows the trajectory of the reaction front for various tensions $U/L$
applied to the system. A first remark would be that the effect of the electric field
on front's motion is {\em very small}. One is also confronted with a counterintuitive
result (that has also been seen experimentally \cite{das90,das91}),
namely that the motion of the front is very slightly {\em slowed down} with
increasing field 
(although the polarity of this field was chosen to favor the reaction).
We had to limit the observation times for these large tensions to $5$ hours, 
because of
the numerical instabilities already invoked above. Of course, for smaller
tensions one can go to longer observation times, like e.g., in 
Figs.~\ref{figure4} and~\ref{figure5}.
 
\begin{figure}[h!]
\begin{center}
\vspace{0.7cm}
\includegraphics[width=\columnwidth]{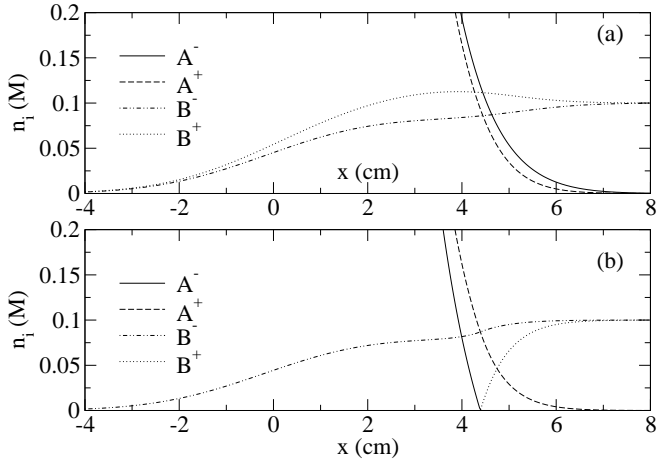}
\end{center}
\caption{Two snapshots of the concentration profiles 
of the ions in the presence
of an applied electric field $U/L=0.5$ V/m:  
(a) without reaction (b) with infinite
reaction rate. The observation time is of $50$ hours and
the values of the other parameters are $L_A=L_B=100$ cm,
$a_0=10$ M, and  $b_0=0.1$ M.}
\label{figure3}
\vspace{0.1cm}
\end{figure}

\begin{figure}[h!]
\begin{center}
\vspace{1cm}
\includegraphics[width=\columnwidth]{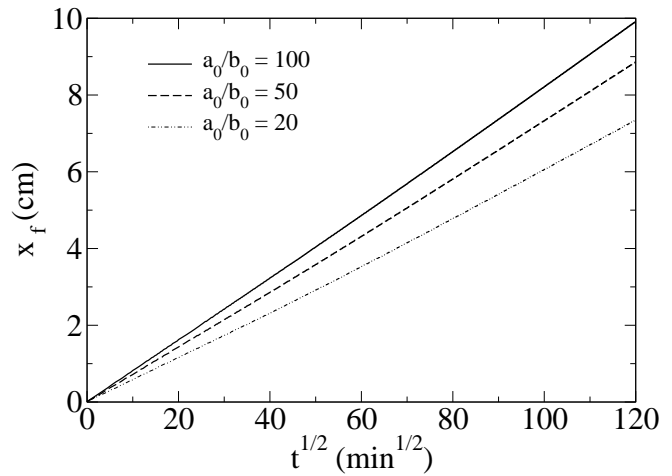}
\end{center}
\caption{Position of the reaction front as a function of the square root of 
time for three
values of the concentrations ratio $(a_0/b_0)$. 
The values of the other parameters
are:  $L_A=1$ cm, $L_B=100$ cm, and $U/L=0.5$ V/m. The observation
time is $10$ days.}
\label{figure4}
\vspace{0.1cm}
\end{figure}

A look at the concentration profiles of the ions, see 
Fig.~\ref{figure3}, allows to sketch
an intuitive explanation of this unexpected behavior. 
Namely, if no reaction takes place,
in the presence of an electric field with a polarity 
$U>0$ the negative $A^-$ ions
will be pushed ahead the positive ones 
(and the corresponding symmetric situation for the $B^{\pm}$ ions),
as it can be seen from the first panel of the figure. 
With an infinite-rate reaction present,
see the second panel of the figure,
the $A^-$ ions are consumed at the front and the $A^+$
ions are forced to move ahead, {\em against the electric field}
(a similar, but symmetric situation occurs for the $B^{\pm}$ ions).
Therefore, the motion should be diffusive 
and has to slow down slightly with increasing field.

However, for all the practical purposes 
(e.g., the production of $C$) one can neglect 
the dependence of the diffusion coefficient 
$D_f$ on the applied tension and consider its fieldless
expression given by \cite{antal99}:
\begin{equation}
\mbox{erf}\left(\sqrt{\frac{D_f}{2D}}\right)=\frac{(a_0/b_0)-1}{(a_0/b_0)+1}\,,
\label{dfront}
\end{equation}
that corresponds to an increase in 
$D_f$ with increasing concentrations ratio $(a_0/b_0)$, see Fig.~\ref{figure4}.
A qualitative comparison of Figs.~\ref{figure2} and \ref{figure4}
shows that $D_f$ is much more sensitive to changes in the concentration ratio
$(a_0/b_0)$ than to changes in the applied field $U/L$.

The finite spatial extent of the system
does not affect the diffusive behavior of the front
until the reaction front reaches the
very vicinity of the borders. Then the front slows down and stops,
and there is a rapid accumulation of $C$ product at the  border. 
However, for the usual observation times we are dealing with, 
this regime of the
front is not relevant,
see Appendix~B for more details.

\subsection{The concentration of the reaction product $C$}

We come now to the most important results from the point of view
of the subsequent pattern formation through precipitation of $C$.
Namely, the study of the influence of the applied 
electric field on the production of $C$ in the
wake of the moving front. 

As well-known (see, e.g., \cite{galfi88,antal99,unger00}), 
in the absence of an applied
electric field the concentration of $C$ left behind the reaction 
front is constant,
and its value $c_0$ is determined by 
the initial concentrations of the ions $a_0$ and
$b_0$, and by their diffusion coefficients. 
In the particular case of equal diffusion coefficients $D$ of the ions,
its value is given by:
\begin{equation}
c_0 \approx  a_0\,K\sqrt{2D/D_f}\,,
\label{c0}
\end{equation}
where $K\equiv (1+b_0/a_0)(2\sqrt{\pi})^{-1}\,\exp(-D_f/2D)$, 
and the diffusion
coefficient $D_f$ of the front is given by Eq.~(\ref{dfront}).

When an electric field is applied to the system, the concentration of $C$
left behind the moving reaction front {\em is no longer constant}, but becomes space-dependent, $c=c(x)$.

There are two factors that determine the instantaneous 
quantity of $C$ that is produced in the wake of the front, namely: 
(i) The velocity of the front, $\dot x_f=d{x}_f(t)/dt$.
As  discussed above,
this one is diffusive, with a diffusion coefficient $D_f$ 
that is essentially
unaffected by the externally applied electric field, 
$\dot x_f=\sqrt{D_f/2t}$.

(ii) The instantaneous value of the current of the reacting 
ions $A^-$ and $B^+$ that arrive at the reaction point,
$|j_{a^-}(x_f(t))|=|j_{b^+}(x_f(t))|$, that has a 
diffusive component, as well as a component proportional to the 
electric current density $j_0(t)$.

\begin{figure}[h!]
\begin{center}
\vspace{1cm}
\includegraphics[width=\columnwidth]{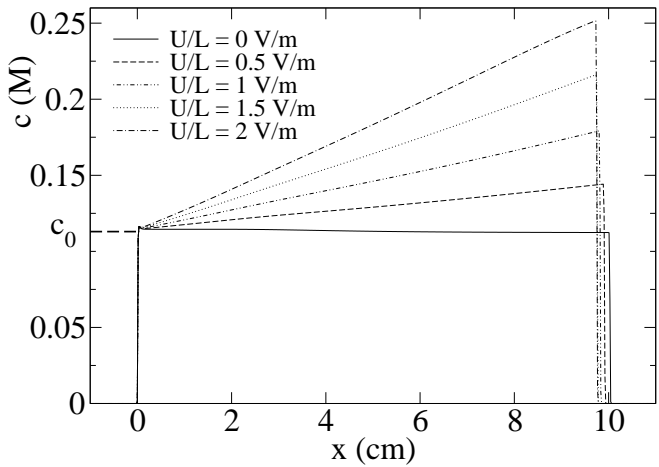}
\end{center}
\caption{The density of the reaction product $C$
left behind the reaction front for different values
of the electric field $U/L$ applied to the system.
The observation time is $t=10$ days.
The values of the other parameters are: $L_A=1$ cm, $L_B=100$ cm, $a_0=10$ M,
and $b_0=0.1$ M.}
\label{figure5}
\vspace{1cm}
\end{figure}

A first regime, that appears without exception 
when the polarity of the applied field favors the
reaction, is a {\em linear increase of $c$ with $x$}, with a 
slope that is proportional with the applied tension,
see Fig.~\ref{figure5}. Moreover, the slope is proportional to
$c_0$, the concentration of $C$ in the absence of the field, given 
by Eq.~(\ref{c0}). 

Thus the numerical results can be summarized in
\begin{equation}
c(x)=c_0[1 + (\alpha_c \,U/L) x]\,,
\end{equation}
and for the profiles in Fig.~\ref{figure5}, one inferres 
$\alpha_c \approx 5 V^{-1}$.

This behavior of the production of $C$ can be easily understood once
we notice that for the relevant observation times the
current density $j_0$
is practically constant (see below),
and thus $|j_{a^-,b^+}(x_f)|$ has a constant $\sim j_0$ and
a diffusive part $\sim 1/\sqrt{t}$.
This fact, in conjunction with the
diffusive motion of the front, leads to a density of 
$C$ in the wake of the front
\begin{equation} 
c(x_f) \sim |j_{a^-,b^+}(x_f)|/\dot{x}_f \;\sim \sqrt{t}+\mbox{const.}
\sim x_f+\mbox{const.}\,,
\end{equation}
i.e., a linear increase with $x$.

The regime of constancy of the current density $j_0$
is discussed in more details in the Appendix~B. It is maintained 
as long as the total electric resistance of the system 
is practically constant.
This means as long as the resistivity of the depletion
(low ion concentrations) region that develops around 
the moving front remains unimportant as compared to the resistance of the
rest `unperturbed' part of the system. Of course, the longer the system, 
the longer this regime lasts. 
For most experimental situations, this is indeed the
only  observable regime.  

\subsection{Reverse polarity of the applied tension}

\begin{figure}[h!]
\begin{center}
\vspace{1cm}
\includegraphics[width=\columnwidth]{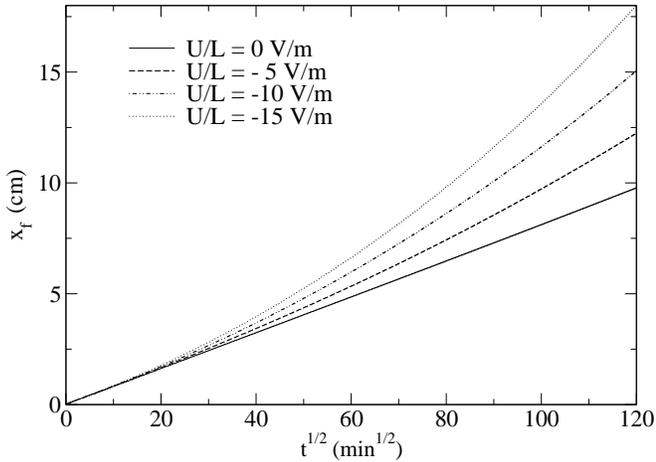}
\end{center}
\caption{The position of the reaction front as a function 
of the square root
of time for different values of the applied field $U/L<0$ 
as compared to the fieldless case. 
One notices the drift component of the motion. 
The observation time is $t=10$ days.
The values of the other parameters are: $L_A=2$ cm, $L_B=40$ cm, $a_0=10$ M,
and $b_0=0.1$ M.}
\label{figure6}
\vspace{1cm}
\end{figure}

We present below the main results 
for a polarity of the applied tension $U<0$
that does not favour the reaction, i.e., tends
to push the $A^-$ and $B^+$ ions far apart.
A first result is that the motion of the 
front is no longer purely diffusive,
but, as can be seen from Fig.~\ref{figure6}, a drift component
emerges. The curves can be well fit
using the expression
\begin{equation}
x_f(t)=\sqrt{2D_ft}+v_f t\,,
\label{drift}
\end{equation}
with, however, the drift part playing 
a significant contribution only for
sufficiently long times and/or high tensions, 
in agreement with experimental data~\cite{sultan00}.

\begin{figure}[h!]
\begin{center}
\vspace{0.7cm}
\includegraphics[width=\columnwidth]{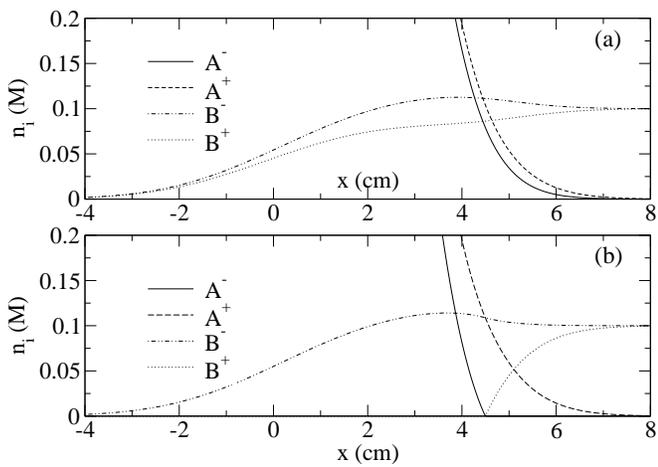}
\end{center}
\caption{Two snapshots of the concentration profiles 
of the ions in the presence
of an applied electric field $U/L=-0.5$ V/m:  
(a) without reaction (b) with infinite
reaction rate. The observation time is of $50$ hours and
the values of the other parameters are $L_A=L_B=100$ cm,
$a_0=10$ M, and  $b_0=0.1$ M.}
\label{figure7}
\vspace{0.1cm}
\end{figure}

A naive attempt to justify
qualitatively this  behavior of the front
appeals to the type of argument
we have already used in Sec.~IIIA for the case of the favorable polarity of $U$ 
(in order to explain, there, the diffusive motion of the front). 
Namely, in the case $U<0$,
the  $A^+$ ions {\em always} move ahead the (reacting) $A^-$'s
(and the symmetric situation
for the $B^{\pm}$ ions), as seen in Fig.~\ref{figure7};
but in this case their motion is in the sense
of the applied field, and might allow for a diffusive, 
as well as a drift
component of front's motion.

\begin{figure}[h!]
\begin{center}
\vspace{1cm}
\includegraphics[width=\columnwidth]{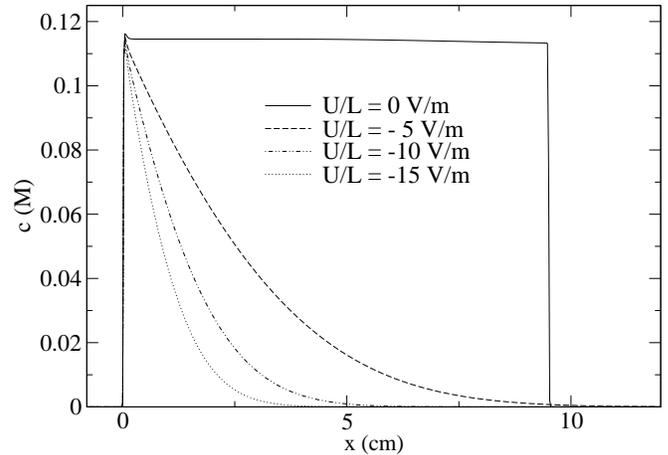}
\end{center}
\caption{The density of the reaction product $C$
left behind the reaction front for 
different values of the applied field $U/L<0$ 
as compared to the fieldless case. 
One notices the rapid extinction of the reaction
with increasing field intensity $|U|/L$.
The observation time is $t=10$ days.
The values of the other parameters are: 
$L_A=2$ cm, $L_B=40$ cm, $a_0=10$ M,
and $b_0=0.1$ M.}
\label{figure8}
\vspace{1cm}
\end{figure}

The concentration $c$ of the reaction product decreases 
nonlinearly with $x$,
as seen in  Fig.~\ref{figure8},
faster when the applied field intensity $|U|/L$ is larger, 
and up to the complete stop of the
reaction (when obviously Eq.~(\ref{drift}) becomes meaningless).
Thus this situation becomes rather rapidly 
uninteresting from the point of view 
of Liesegang pattern formation, as it leads
to the disappearance of the pattern.

\section{Conclusions and perspectives}

In this paper we examined the influence of an applied electric field on the
propagation of the reaction front and on the production of $C$
for an $A^-+B^+ \rightarrow C$ reaction in a one-dimensional geometry that is
specific to Liesegang-pattern experiments. We deliberately concentrated 
the presentation of our results on the experimentally relevant parameters and
observation times (although, for other ranges of the parameters and
longer observation times other regimes might come into play). Our main
conclusions are that for a polarity of the 
applied field that favours the chemical
reaction, (i) the motion of the reaction front is diffusive, with a diffusion
coefficient practically unaffected by the applied field, and
(ii) the concentration of the reaction product $c(x)$ left behind the front
increses linearly with $x$, with a slope that is proportional to the applied
tension. 

For the reverse polarity (i.e., that unfavours the reaction), (i) the motion of the front
has a diffusive as
well as a drift component, and (ii) the concentration $c(x)$
of the reaction
product left by the front decreases nonlinearly with $x$
up to the complete stop of the reaction.

The above results on the motion of the reaction front 
are in agreement with the
experiments in~\cite{das90,das91,sultan00}, but appear to be at variance with
the front motion in~\cite{lagzi02}.

The data we presented are based on a set of approximations.
Any of them can be
relaxed and/or improved, but we have evidences that none of them affects
essentially our conclusions.

The next step would be to use the above results to determine the influence of 
the applied tension on the characteristics of the Liesegang patterns. This can
be easily done in the framework of the spinodal decomposition scenario. The
dynamics of the $C$ particles is still described by a Cahn-Hilliard equation,
but this time the source term is modified (as compared to the fieldless case,
see \cite{antal99}) in agreement with our conclusions (i) and (ii) above. A
rapid examination of these elements already allows us to predict for a 
favourable polarity
$U>0$ a decrease in
the spacing between the bands (the more pronounced the higher the applied
tension), and, for sufficiently long times and/or high enough tensions, the
disappearance of the pattern. 
For the reverse polarity $U<0$, the pattern (if any) will present 
an increase in the spacing between bands as compared to 
the fieldless case, followed by a
rather rapid disappearance of the pattern.
A detailed study of this problem will be given in a further publication.

\appendix

\section{The general one-dimensional model}

We present below the general one-dimensional evolution equations for the
concentration profiles of the participating species when we do not the 
approximations of  ideally strong acid and basis,  infinite reaction
rate, and  equal diffusion coefficients of the ions (see the main text).

When relaxing the assumption of ideally strong acid and basis for
the electrolytes $A$ and $B$, one has to take into account their diffusive
motion and their finite dissociation, i.e., the following dynamics of their
concentration profiles $a(x,t)$ and $b(x,t)$:
\begin{eqnarray}
\frac{\partial a(x,t)}{\partial t}&=&D_a\,\frac{\partial^2 a}{\partial x^2}
-\lambda_a(K_aa-a^-a^+)\,,
\label{ab-ionprod1}\\
\frac{\partial b(x,t)}{\partial t}&=&D_b\,\frac{\partial^2 b}{\partial x^2} -\lambda_b(K_b\,b-b^-b^+)\,.
\label{ab-ionprod2}
\end{eqnarray}
Here $a^-(x,t)$, $a^+(x,t)$, $b^+(x,t)$, and $b^-(x,t)$ 
are the concentrations of the appropriate
ions resulting from $A$ and $B$, respectively,
and  $\lambda_{a,b}$ are the relaxation
constants to the respective dissociation
equilibria characterized by the dissociation constants $K_a$ and $K_b$.

As already mentioned, $A^+$ and $B^-$ are not reacting, while
the ions $A^-$ and $B^+$ are reacting {\em irreversibly}
with a certain reaction rate $k$, that we shall consider {\em finite},
\begin{equation}
A^-+B^+ \rightarrow C\,. 
\label{a-b+reaction}
\end{equation} 
We assume that the dynamics of the {\em inert} reaction product 
$C$ has no feedback on the dynamics of the reagents \cite{footnote1}. 

The macroscopic evolution equations 
for the concentrations of the ions
in the presence of an electric field $E(x,t)$, by considering 
{\em different} diffusion coeficients of the ions, read:
\begin{eqnarray}
\frac{\partial a^-(x,t)}{\partial t}&=&D_a^-\frac{\partial^2 a^-}{\partial x^2}+
\lambda_a(K_aa-a^-a^+) -ka^-b^+ \nonumber\\
&&-D_a^-z_a^-\,\frac{F}{RT}\frac{\partial (a^-{E})}{\partial x}\,,
\label{electroneut1}\\
\frac{\partial b^+(x,t)}{\partial t}&=&D_b^+\frac{\partial^2 b^+}{\partial x^2}+
\lambda_b(K_bb-b^+b^-) -ka^-b^+\nonumber\\
&&-D_b^+z_b^+\,\frac{F}{RT}\frac{\partial (b^+{E})}{\partial x}\,,
\label{electroneut2}\\
\frac{\partial a^+(x,t)}{\partial t}&=&D_a^+\frac{\partial^2 a^+}{\partial x^2}+
\lambda_a(K_aa-a^-a^+)\nonumber\\
&&-D_a^+z_a^+\,\frac{F}{RT}\frac{\partial (a^+ {E})}{\partial x}\,,
\label{electroneut3}\\
\frac{\partial b^-(x,t)}{\partial t}&=&D_b^-\frac{\partial^2 b^-}{\partial x^2}+
\lambda_b(K_bb-b^+b^-)\nonumber\\
&&-D_b^-z_b^-\,\frac{F}{RT}\frac{\partial (b^-{E})}{\partial x}\,.
\label{electroneut4}
\end{eqnarray}
Here $D_{a,b}^{\pm}$ are the respective
diffusion coefficients of the ions, $F=qN_A$ is  Faraday's constant 
(i.e., the electric charge transported by a mole
of monovalent positive ions), $R$ is the universal gas constant, 
while $T$ is the temperature.
The $z_i$-s are signed
integers giving the charge of the $i$-th ion, 
$i=a^{\pm},b^{\pm}$, in units of elementary charge $q$, and we hold
the {\em local electroneutrality} assumption \cite{rubinstein90}, i.e.,
$$\sum\limits_i z_i n_i(x,t)=0$$ 
at any point and at any time 
($n_i(x,t)$ are the concentrations of the ions).
It is useful to recall that indeed local electroneutrality
assumption is well-justified for Liesegang type of experiments: at
characteristic ion concentrations $\sim 10^{-3}$ -- $10$ M
usually present in Liesegang experiments, the Debye screening length 
is of the order of $\sim 10^{-10}$ m -- $10^{-8}$ m; thus, it is
indeed negligible as compared to
the other length scales present in the system -- 
lengths of the precipitation zones of the order 
$\sim 10^{-3}$ m, and width of the reaction zone $\sim 10^{-6}$ m. 
See \cite{unger00,unger99} for a more detailed discussion of this point.

As known from textbooks \cite{rubinstein90},
the  local electric field $E(x,t)$ 
in the above equations is determined both by the externally applied field
and by the local electroneutrality condition, and it is given by:
\begin{equation}
{E(x,t)}=
\frac{\displaystyle\frac{{j}_0(t)}{q}+\sum\limits_iD_iz_i
\displaystyle\frac{\partial n_i}{\partial x}}
{\displaystyle\frac{F}{RT}\,\sum\limits_iD_iz_i^2n_i}.
\label{Efield}
\end{equation}
${j}_0(t)$ is the electric  current density,
flowing through  the system. Note that in view of the electroneutrality
condition ${j}_0$ is divergence-free, i.e., for the one-dimensional case
it is only time-dependent.

Suppose  that a {\em constant} voltage difference $U=V(L_B)-V(-L_A)$
is applied between the two ends of the system.
Then, according to Eq.~(\ref{Efield}),
\begin{equation}
\int\limits_{-L_A}^{L_B}dx{E(x,t)}=
\int\limits_{-L_A}^{L_B}dx\frac{\displaystyle\frac{j_0(t)}{q}+\sum\limits_iD_iz_i
\displaystyle\frac{\partial n_i}{\partial x}}
{\displaystyle\frac{F}{RT}\sum\limits_iD_iz_i^2n_i}=-U\,.
\label{Efieldint}
\end{equation}
Thus the instantaneous value of the  current density $j_0(t)$ is
determined by the applied tension and by the instantaneous concentration
fields $n_i(x,t)$:
\begin{equation}
\frac{j_0(t)}{q}=\frac{-\displaystyle\frac{FU}{RT}-
\displaystyle\int_{-L_A}^{L_B}dx \sum\limits_iD_iz_i
\displaystyle\frac{\partial n_i}{\partial x}
\left(\sum\limits_iD_iz_i^2n_i\right)^{-1}}
{\displaystyle\int_{-L_A}^{L_B}dx \left(\sum\limits_iD_iz_i^2n_i\right)^{-1}}\,.
\label{J0}
\end{equation}
 At its turn, $j_0(t)$ 
determines through Eq.~(\ref{Efield}) the 
instantaneous electric field ${E}(x,t)$, 
that enters the evolution equations~(\ref{electroneut1})--(\ref{electroneut4}).

One concludes that
Eqs.~(\ref{ab-ionprod1}), (\ref{ab-ionprod2}), (\ref{electroneut1}),
(\ref{electroneut2}), (\ref{electroneut3}), (\ref{electroneut4}) for the
concentration profiles,
(\ref{Efield}) for the electric field, and (\ref{J0}) for the 
current density are all coupled in a highly nonlinear, intricate way.
In our actual study (see Sec.~II), we used  several simplifications
to make these equations  more tractable.

\section{More numerical results}

Here we shall present additional
results from our numerical simulations. We did
not  describe them in the main
text, in order to make the essential results more transparent.
However, we think they may put
supplementary light on our conclusions, as
well as on some controversial points of
previously published results.

\subsection{Influence of finite-size effects on front's motion}

This section intends to strengthen the conclusion presented in the main text,
namely that the finite-size effects do not influence the motion 
of the front as
long as this one does not `hit' the right border.
This reinforce also the idea that the factors responsible 
of front's motion are
{\em strictly local}.

\begin{figure}[htb]
\begin{center}
\vspace{1cm}
\includegraphics[width=\columnwidth]{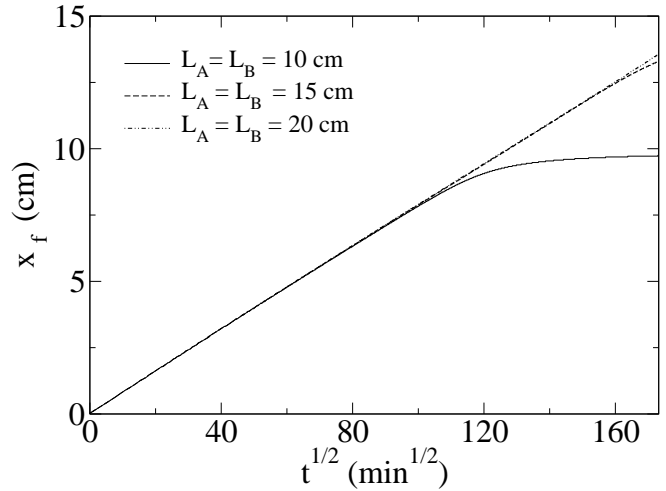}
\end{center}
\caption{Position of the reaction front as a function of the square root of 
time for different
values of the system length.
The values of the other parameters are:
$a_0=10$ M,  $b_0=0.1$ M, 
and applied electric field $U/L= 0.5$ V/m.
The observation
time is of $500$ hours.}
\label{figure9}
\vspace{0.1cm}
\end{figure}

 In Fig.~\ref{figure9}
we considered three system sizes and observed the long-time behavior: 
(i) for the smallest system the front hits the right border 
during the simulation time;  
(ii) for the intermediate system the front starts to `feel' 
the border at the end of the simulations;
(iii) in the the longest system, the front is unaffected by the borders during
the simulation time. 
 
Note however the very long time required to see such border
effects even for the shortest system 
(about $10^{4}$ minutes  for a system with $L_B$ of only $10$ cm). 
In most  of the real experiments
(for which both the system length is bigger -- 
of at least a few tenths centimeters --
and the observation time is usually smaller -- a few days),
as well as in most of our simulations, 
this slowing-down regime of the front is practically {\em never attained}.

\subsection{The current density $j_0$}

For a constant applied tension $U$
the current density $j_0$ is a function of time only;
it is a {\em global} quantity, being determined 
by an integral over the entire system involving the instantaneous 
concentration fields. Consequently, $j_0(t)$ is sensitive to finite-size
effects all along its evolution. The different regimes of $j_0(t)$ result 
essentially from the interplay between the {\em evolving} relative electric
resistances of the {\em depletion zone} 
and that of the rest, `unperturbed part' of the system.

The depletion zone is
defined as the region where the concentrations of the ions are
significantly smaller than their initial values. 
This region expands progressively, more
or less rapidly (depending on the applied tension)
around the moving reaction front.
It can develop, of course, up to the size that 
is allowed by the borders of the
system. It is a region of higher resistivity than the 
`unperturbed' part of the system.
There are two elements (with distinct temporal evolution) that determine the
electric resistance of the depletion zone, namely (i) its spatial extent and
(ii) its degree of depletion, i.e., the concentrations of the ions 
in the region.
Note that the resistivity of the unperturbed part of the system is determined
by its spatial extent, which equals $L$ minus the
length of the depletion zone.

\begin{figure}[h!]
\begin{center}
\vspace{1cm}
\includegraphics[width=\columnwidth]{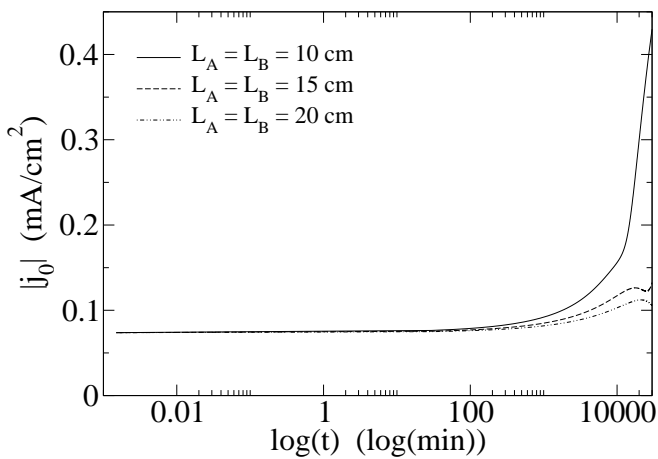}
\end{center}
\caption{The absolute value of the current density $|j_0|$ 
as a function of time (logarithmic scale)
for various system sizes -- the same systems as those in Fig.~\ref{figure9}.
The values of the other parameters are:
$a_0=10$ M, $b_0=0.1$ M,
and applied electric field $U/L= 0.5$ V/m.
The observation
time is of $500$ hours.}
\label{figure10}
\vspace{0.1cm}
\end{figure}

At the initial stages, the spatial extent of the depletion zone
is small as compared to the rest of the system, and the 
concentrations of the ions are not extremely low yet.
Therefore, its contribution to the electrical resistance of the column is
negligible; during this stage, the resistance of the column is essentially
constant and determined by the `unperturbed' part of the system. 
Consequently, one obtains an
initial {\em regime of constant current density}. 
This can be observed in Fig.~\ref{figure10}, 
from which one also realizes that this is
actually the regime that is {\em accesible experimentally}, 
and thus the only of interest to us.

Although we shall not concentrate on them further
(as not relevant for the typical experimental domains) 
let us briefly describe the other regimes that can be
seen in  Fig.~\ref{figure10}, in the final part of the simulations. 

(i) For the shortest system, the depletion zone 
cannot develop significantly before
the front reaches the right border. 
Then the current passes directly from the
constant-value regime to a rapid increase.
When the front is in the vicinity of the border, the
right-hand (with respect to the front) high-resistivity region of  
small concentration of the ions reduces progressively in size, 
while there is an advancement, behind the front, 
of a region of low-resistivity, high ion concentrations.
This generates a decrease in the resistance of the
system, and thus the observed increase in the current density.

(ii) For the intermediate and the longest system, the depletion region 
has more time to develop. At a certain time (shorter for the intermediate
system), the electric resistivity of the depletion zone becomes comparable to
that of the rest of the system. Then a transient regime starts -- seen as an
increase in $|j_0|$ for both systems. Finally, the
increasing resistivity of the depletion zone becomes dominating in the system, 
causing a monotonous decrease in the current density $|j_0|$.
However, the intermediate system does not `have time' 
to develop this regime further: it hits the border
and  starts the corresponding increase in $|j_0|$, according to the 
mechanism sketched at point (i) above. A longer time simulation 
(not shown here) for
the longer system shows a continuation of this regime, with 
$j_0(t) \sim 1/\sqrt{t}$.

\subsection{The electric field}

\begin{figure}[h!]
\begin{center}
\vspace{1cm}
\includegraphics[width=\columnwidth]{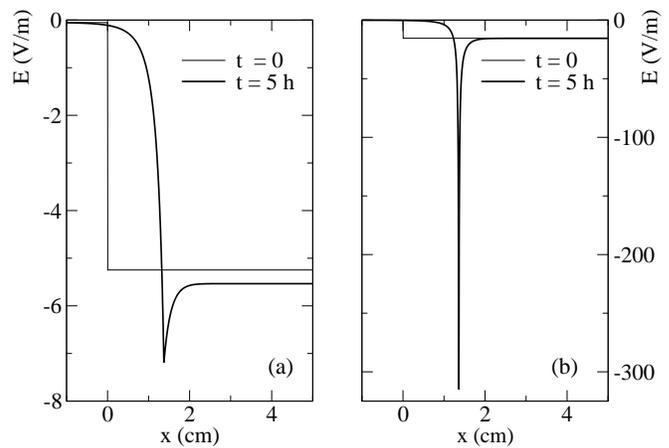}
\end{center}
\caption{Snapshots of the electric field $E(x,\,t)$ at two times and for
two different values of the electric field $U/L$ applied to the system: 
(a) $U/L$ = 5 V/m (b) $U/L$ = 15 V/m. 
The values of the other parameters
are: $L_A=1$ cm, 
$L_B=20$ cm, $a_0=10$ M, and $b_0=0.1$ M.}
\label{figure11}
\vspace{0.1cm}
\end{figure}

The electric field is not constant along the system, as can be seen from
Fig.~\ref{figure11}, 
in contradiction with the fundamental assumption in
\cite{feeney83}. This can be easily understood by 
inspection of Eq.~(\ref{Efield}), taking into account that
the concentration fields vary along the system.
For a polarity of the externally applied tension $U$
that favors the reaction, 
the  absolute value of the electric field
increases rapidly in the vicinity of
the moving front.  This effect is due to the decrease in both the 
the concentrations of the reacting ions $A^-$ and $B^+$ (due to the reaction),
and the concentrations of the background ions 
$A^+$ and $B^-$ (because the polarity of the electric field pushes them away
from the reaction zone). As a result, the sum in the denominator of the
expression~(\ref{Efield}) determining the electric field
$\sum_i D_i z_i^2 n_i$ becomes small in the vicinity of the reaction front, 
and thus the local field becomes very large.

This rapid -- both {\em spatial and
temporal}, see Fig.~\ref{figure11}
-- increase in the electric field
renders the numerical procedure
unstable after some time. The other element that contributes to this
instability is also connected with the small densities of the ions around the
reaction zone: namely, from one integration step to another, the local
concentration profiles may acquire unphysical negative values. These
instabilities develop faster for larger applied tensions; 
their appearance can be 
delayed by decreasing both the spatial and temporal discretization
steps. One is thus limited in the choice of the
upper value of the applied tension $U$ by a `reasonable' choice of the spatial
and temporal discretization steps, that allow the study of the system for 
time intervals that are physically relevant.
From a technical point of view, these instabilities are promptly notified by the
value of the electric field at the reaction point, 
that acquires an unphysical temporal
evolution at the moment of the onset of the instabilities.

The appearance of these numerical instabilities  can be also postponed by: 
(i) giving up the infinite dissociation rate
approximation, and allowing the $A$ and $B$ electrolytes
to diffuse towards the reaction zone and to supply it with ions $A^{\pm}$ and
$B^{\pm}$ through dissociation;
(ii) taking into account the finiteness of the reaction rate $k$ between 
$A^-$ and $B^+$, that leads to a slower decrease of the concentration of the
reacting ions in the
reaction zone. Note that in this case the reaction zone acquires 
a finite spatial extent (i.e., it is
no longer reduced to a single point).

Finally, we should mention in connection with the electric field that,
in real systems, this one has supplementary inhomogeneities
related to the geometry of the system and, furthermore, the electric
stability of
the gel column may also reduced by electric effects at the walls of the
container~\cite{lagzi04}.

\subsection{The concentrations of the ions}

The concentration profiles $n_i(x,\,t)$ of the ions have a 
very complex evolution, that is
determined by their diffusion coefficients, the initial concentrations,
the length of the system, and the applied electric tension. 
A systematic presentation of all the
corresponding effects would be tedious,
and thus we shall present below only a few
relevant elements.

\begin{figure}[h!]
\begin{center}
\vspace{1cm}
\includegraphics[width=\columnwidth]{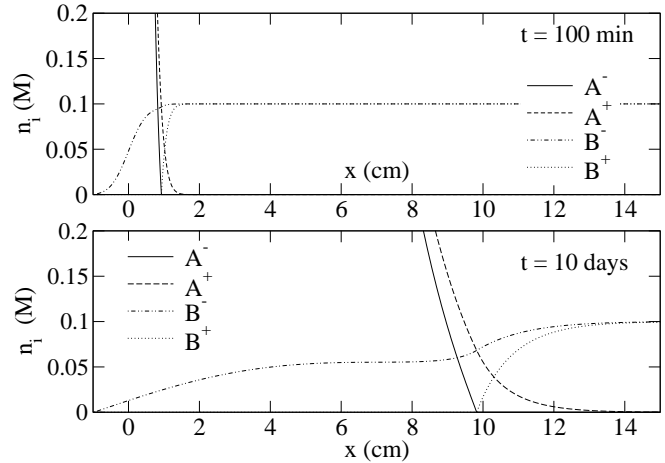}
\end{center}
\caption{Two snapshots of the concentration profiles 
of the ions in the vicinity of the
reaction front at two different times. 
The values of the parameters of the system
are: $L_A=1$ cm, 
$L_B=100$ cm, $a_0=10$ M, and $b_0=0.1$ M, 
and applied electric field $U/L = 1$ V/m.}
\label{figure12}
\vspace{0.1cm}
\end{figure}

For a given system (i.e., with fixed parameters), a qualitative 
typical temporal evolution  of the concentration
profiles corresponds to a reduction of the concentration 
of the ions in the vicinity of
the moving reaction front, 
as well as to a progressive extension of this depletion zone
(which, recall, is defined as the region where the concentrations 
of the ions are
significantly smaller than their initial values); of course,  
as long as permitted by the boundaries,
on which the concentrations of the ions are fixed. 
This is illustrated in Fig.~\ref{figure12}.

This depletion effect is more pronounced when the polarity of the applied
tension is such that it favors the reaction than in the case of the opposite
polarity. This can be easily understood: 
in an electric field that favors the reaction, 
not only the disappearance of
the ions $A^-$ and $B^+$  is  favorized through reaction, 
but also the background ions $A^+$
and $B^-$ are pushed away from the reaction zone. Of course, in a
field of reverse polarity, the nonreacting ions $A^+$ and $B^-$ 
are pushed towards each other, and
the active ions $A^-$ and $B^+$ are 
pushed away from
each other (and thus their consumption through reaction is reduced;
this leads, in the end, to an extinction of the reaction). 

Finally, for a polarity of the applied tension that is favorable 
to the reaction,
the expansion of the depletion zone is more rapid and pronounced 
for larger tensions.
This determines the local increase of the 
electric field, and, in the long time, the numerical 
instability of the integration procedure
(as discussed in the previous subsection). 

\subsection{The concentration of the reaction product $C$}

\begin{figure}[h!]
\begin{center}
\vspace{1cm}
\includegraphics[width=\columnwidth]{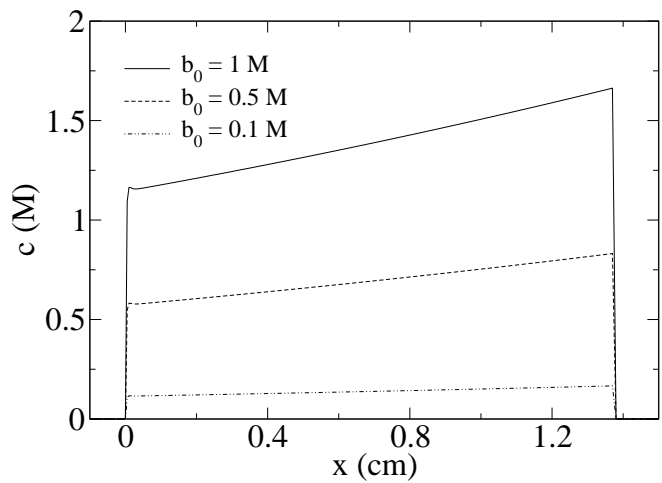}
\end{center}
\caption{The density of the reaction product $C$ left behind the reaction front
for different values of the concentration $b_0$. 
The observation time is $t = 5$ hours.
The values of the parameters of the system
are: $L_A=1$ cm, 
$L_B=20$ cm, fixed ratio $a_0/b_0=100$, and the applied electric field 
$U/L$ = 5 V/m.}
\label{figure13}
\vspace{0.1cm}
\end{figure}

For the regime of constant $j_0$, as discussed in the main text, the
concentration of the reaction product $c(x)$ increases linearly with $x$ with a
slope proportional to the fieldless value $c_0$, as illustrated by
Fig.~\ref{figure13}. 

This initial behavior of the production of $C$ is determined by and 
lasts as long as the initial regime of
constancy of the current density $j_0$, and therefore 
{\em it is the only regime  
on  which we  focus in our study}.
However, as discussed above in the corresponding subsection, 
the current density $j_0$ may develop other regimes 
(if allowed by the size of the system). Then, of course, these regimes will 
be reflected by the production of $C$.
In particular, the transient increase (if any)
in $|j_0|$ between the initial regime of constancy
and the regime $j_0 \sim 1/\sqrt{t}$ is reflected by a transient 
increase in the production
of $C$ (a ``bump" in the profile $c(x)$). The regime $j_0 \sim 1/\sqrt{t}$
itself leads to a constant deposition of $C$ in the wake of the front;
the plateau value of $c$ 
varies with $a_0$ and $b_0$ (much in the same way as $c_0$), 
but is rather insensitive to the applied tension $U$.
Finally, when the reaction front `hits' the right border, there is a great
increase in the production of $C$ at the border.

\acknowledgments

We thank I. Lagzi for very useful discussions.
This research has been partly supported by the 
Swiss National Science Foundation and
by the Hungarian Academy
of Sciences (Grants No.\ OTKA T043734 and TS 044839).



\begin{thebibliography}{}
\bibitem{footnote2} Thus the problem of the dependence of the width of the
reaction front on the applied electric field is not approached here, 
but left for further studies.
\bibitem{footnote3} $U=$ constant is the most common
experimental situation. But one can equally well consider experimental
situations for which the current density $j_0$ is maintained constant.
However, as discussed in some more detail in 
Appendix~B2, for the experimentally relevant 
observation times there is no
significant difference between these two situations.
\end{thebibliography}
\end{document}